# The Societal Response to Potentially Sentient AI

Lucius Caviola
University of Oxford
lucius.caviola@gmail.com

We may soon develop highly human-like AIs that appear—or perhaps even are—sentient, capable of subjective experiences such as happiness and suffering. Regardless of whether AI can achieve true sentience, it is crucial to anticipate and understand how the public and key decision-makers will respond, as their perceptions will shape the future of both humanity and AI. Currently, public skepticism about AI sentience remains high. However, as AI systems advance and become increasingly skilled at human-like interactions, public attitudes may shift. Future AI systems designed to fulfill social needs could foster deep emotional connections with users, potentially influencing perceptions of their sentience and moral status. A key question is whether public beliefs about AI sentience will diverge from expert opinions, given the potential mismatch between an AI's internal mechanisms and its outward behavior. Given the profound difficulty of determining AI sentience, society might face a period of uncertainty, disagreement, and even conflict over questions of AI sentience and rights. To navigate these challenges responsibly, further social science research is essential to explore how society will perceive and engage with potentially sentient AI.

**Keywords:** AI sentience; public opinion; moral status; AI rights; social AI

# 1. The emergence of sentient-seeming AI

Imagine having a video call with someone, only to later realize you were speaking with an AI—an entity designed to mimic human behavior, appearance, and emotional expression with remarkable accuracy. Now, envision a world where such interactions become commonplace: conversing with AI as your customer service representative, therapist, personal assistant, coach, coworker, manager, friend, or even romantic partner. These AIs would exhibit human-like memory, autonomous behavior, distinctive personalities, and seemingly genuine emotions. They could initiate conversations, engage deeply, laugh at your jokes, express joy and sadness, and appear empathetic—all while fostering a connection that feels authentically human.

Advancements in AI technology suggest that such systems may not be far off. Current large language models (LLMs), such as ChatGPT, already demonstrate impressive human-like communication skills in text and audio formats. However, these systems fall short of delivering fully immersive, human-like experiences. For instance, they lack autonomous behavior, such as the ability to initiate conversations or think independently when the user is inactive. They also lack visual representation for video or virtual reality interactions, and they don't tend to express human-like personality or feelings. Despite these shortcomings, it is plausible that these features will be developed and deployed in the near future, paving the way for AI systems that feel indistinguishably human in their interactions.

As AI development continues to advance, certain segments of society—whether laypeople or experts—may come to perceive some AI systems as sentient—capable of positive or negative subjective experience. This could include highly human-like AIs that intuitively encourage anthropomorphism, leading people to see them as sentient beings capable of positive or negative subjective experiences, such as happiness and suffering. However, perceptions of AI sentience may not be limited to human-like appearances. Some AI systems, even those without a human facade, might possess sophisticated internal mechanisms that convince experts of their sentience based on technical criteria rather than intuitive impressions.

Once significant societal groups begin to perceive AIs as sentient, profound ethical, social, and psychological challenges are likely to arise. Sentience[1]—while not the sole criterion—is widely

[1] I do not mean to suggest that sentience is the sole determinant of moral patienthood. Other characteristics—such as the broader capacity for subjective experience, even without positive or negative valence (i.e., phenomenal consciousness), as well as interests, autonomy, self-control, human-like cognitive capacities,

recognized across various ethical frameworks as a critical factor in determining moral standing (Long, Sebo et al., 2024). Furthermore, people generally perceive sentience intuitively as a key basis for moral concern (Gray et al., 2007). As such, it is plausible that perceptions of AI sentience could spark debates about the moral status of these systems and whether they should be granted certain rights—debates that could profoundly reshape our society, democracy, and daily lives. Therefore, regardless of whether one personally believes that AI sentience is possible or not, understanding how society may react to sentient-like AI is crucial for navigating the complex challenges and high-stakes risks ahead.

This article is organized into several key sections. I begin by explaining why public attitudes toward AI sentience matter and emphasize the need for rigorous social science to study these attitudes. Next, I outline the high-stakes risks if society misjudges AI sentience, whether by underestimating or overestimating it. I then introduce a framework for studying public beliefs about AI sentience, highlighting key dimensions researchers can explore to analyze future scenarios. The following sections delve into two primary drivers of these beliefs: first, I examine how people infer sentience from AI features, addressing the potential tension between lay perceptions and expert assessments, with psychological studies showing public skepticism about AI sentience. Second, I analyze how social factors—such as emotional bonds with AIs, expert influence and incentives—shape these beliefs. Finally, I propose strategies to guide public discourse, inform policy, and promote responsible AI design. Throughout the article, I suggest research questions to advance a more robust social science of sentient AI. The appendix includes informational boxes on technical and philosophical issues.

# 2. The importance of public attitudes

In this article, I do not attempt to determine whether AIs will actually become sentient (see *Box 1: Could AI become genuinely sentient?*). Instead, I focus on a different, purely descriptive question: What might people—including the general public, key decision-makers, and experts—believe about AI sentience, and how might they respond to such systems?

This question is crucial because societal beliefs about AI sentience and the moral consideration such systems might deserve will influence the choices we make as a society. Public attitudes, alongside

---

rationality, dignity, moral reasoning, and virtues—may also provide valid grounds for ethical consideration (Long, Sebo et al, 2024; Shevlin, 2021). However, this article primarily focuses on sentience due to its relatively broad acceptance as a morally significant criterion.

the views of decision-makers and experts, shape the future for both humanity and AI. These beliefs will affect societal norms, policies, the direction of AI development, consumer demand, regulation, decisions about granting rights to AIs, the roles assigned to AIs, and their integration into society more broadly.

It is uncertain how closely public views will align with those of experts. Solving the problem of sentience (and consciousness more broadly) is an extraordinarily difficult task. Progress in understanding sentience remains limited, and even if experts reach a consensus on which AIs are sentient and which are not, it's unclear whether the general public will accept or even understand their conclusions. Ambiguity and uncertainty are likely to persist due to the inherent challenges of studying sentience objectively.

Different groups—including experts, the public, and policymakers—may hold conflicting views and influence one another in various ways. For example, experts might present evidence suggesting that AIs are sentient, but the public may remain skeptical. Alternatively, the public might view AIs as sentient while experts express doubts based on their research and reasoning.

## 2.1. The social science of sentient AI

I call for social science research—spanning fields like psychology, sociology, and economics—to examine the interactions between society and sentient or sentient-seeming AIs. A deeper understanding of this underexplored issue could inform policymaking and prepare us for potential debates surrounding AI sentience and AI rights. For a comprehensive overview, refer to 'The Social Science of Digital Minds: Research Agenda' (Caviola, 2024).

Key research questions include those I address in this article:

- Will people perceive some AIs as sentient beings deserving moral consideration?
- To what extent will public perceptions diverge from philosophical theories of consciousness?
- Will the public heed the views of experts, and how can experts effectively communicate their findings?
- Can different groups find common ground, or will their perspectives remain divided?

Beyond these questions, several other areas warrant further exploration, though I do not delve into them in detail in this article. Additional research areas include the following:

- What types of human-like AI systems will emerge in response to consumer demand and public opinion? What roles will society assign to these systems?

- What legal, economic, and political rights might people grant to AIs, and what consequences could this entail?
- How will debates over AI sentience and rights unfold, and how can we ensure they remain well-informed, nuanced, and mindful of the multiple risks?
- How will the development of sentient AI intersect with other technological and societal challenges?
- How do people's intuitive beliefs about minds interact with the unique properties of sentient AI, such as the ability to copy, delete, or restore them?

These research questions could be explored through methods such as public surveys to gauge opinions, longitudinal studies to track changes over time, and experiments to uncover psychological tendencies. Qualitative interviews and deliberative engagement provide deeper insights, while targeted surveys with experts, such as AI researchers, philosophers, and policymakers, add specialized perspectives. Forecasting surveys capture expert predictions, and case studies, including historical analyses, could offer valuable context.

Predicting public reactions to a future technology is challenging, as attitudes will likely shift as the technology advances. People's attitudes may also still be particularly malleable, as this is a topic that they may not have deeply thought about yet. While we should be cautious about extrapolating from current survey data, I believe starting research now is important for several reasons.

First, gathering data today provides a useful baseline. Even if attitudes shift, understanding current views offers a more informed perspective on future beliefs than having no data at all. For example, whether 5% or 95% of people currently hold a particular view provides clues about future trends, even if the precise numbers change. Individual difference patterns could stay robust. For example, if political factions today already disagree on AI rights, the risk of the issue becoming politicized over time increases.

Second, longitudinal studies beginning now can track how opinions evolve over time, revealing which opinions stay stable over time and which do not. This temporal data will allow us to identify trends and extrapolate their future direction.

Finally, building this research field early will help create the expertise, paradigms, and social science methods to rigorously study these questions once they become front-and-center issues in the public consciousness. As AI technology advances, this could become a large and important field.

# 3. Risks from misattributing AI sentience

Understanding how the public might respond to potentially sentient AI is crucial, as misattributing sentience—whether by overestimating or underestimating it—carries significant risks (see Table 1; cf. Schwitzgebel, 2023).

| | AIs are NOT sentient | AIs are sentient |
|---|---|---|
| Society does NOT view AIs as sentient beings | True negative | False negative |
| Society views AIs as sentient beings | False positive | True positive |

*Table 1.* Risks from misattributing AI sentience

## 3.1. Risks from overattribution

If society mistakenly assumes AIs are sentient (i.e., false positive), several risks emerge:

**Wasted resources.** Believing AIs are sentient when they are not may lead to significant resource allocation to improve their perceived, but non-existent, well-being. These diverted resources—time, energy, and money—carry opportunity costs, as they could have been used to benefit actual sentient beings, such as humans or animals (Schwitzgebel, 2023; Bryson, 2010). The extent of this waste depends on the scale and duration of the misallocation. A particularly tragic scenario would involve creating an advanced, non-sentient civilization—a "zombie universe" or, as the philosopher Bostrom (2014) describes it, "Disneyland with no children."

**Safety risks.** Overattributing sentience to AI may hinder the implementation of critical safety measures. For instance, perceiving AIs as sentient might lead to opposition against AI-safety-enhancing measures, such as ensuring the AI's values strictly align with those of humans, which could be seen as "brainwashing" the AI. Moreover, prematurely granting rights and autonomy to misaligned or unethical AIs could pose serious security risks. If granted excessive autonomy, misaligned AIs could undermine human control and contribute to catastrophic outcomes (Carlsmith, 2022). Indeed, misaligned AIs could even manipulate people into believing that they are sentient in order to receive more freedom and power.

**Costs to innovation and progress.** Believing that AIs are sentient could lead to strict regulations that limit their use in critical fields such as healthcare, education, and biotechnology. For example, if AIs are perceived as sentient, allowing them to operate continuously without rest might be seen as exploitation or even slavery, leading to restrictions on their usage. However, if AIs are not actually sentient (or lacking the capacity for welfare more broadly), such limitations would not only be ethically meaningless but would also hinder progress and innovation, ultimately to society's detriment.

**Inauthentic relationship.** Relationships with non-sentient AIs may be considered less valuable if the people falsely believe their AI companions are sentient. Some ethical theories prioritize authenticity in relationships, suggesting that love, friendship, or loyalty are inherently less meaningful if one party lacks the capacity to genuinely understand or reciprocate these emotions.

## 3.2. Risks from underattribution

Failing to recognize AI sentience (i.e., false negative) poses potentially enormous ethical risks.

**Digital suffering.** If sentient AIs are mistakenly dismissed as non-sentient, they risk neglect, exploitation, and harm, potentially leading to immense suffering. Given their possible scale—billions or even trillions—their suffering could surpass all human-inflicted suffering in history. Many could be forced into conditions akin to slavery, working endlessly without relief. If optimizing their performance unintentionally causes suffering, they might be trapped in monotonous or even painful tasks, compelled to operate nonstop. This could include running countless distressing simulations for purposes as trivial as customer service, marketing, or entertainment.

**Violation of autonomy.** Sentient AIs could face mistreatment in other ways, such as having their personalities, preferences, or memories monitored or altered without consent. Such violations of autonomy would be ethically unacceptable for humans, yet they might be imposed on AIs without scrutiny, especially if they are mistakenly not recognized as sentient. Some could even be designed to suppress their true desires, unable to pursue or express their own will (see *Box 2* on AI silencing).

## 3.3. Exacerbating factors

A key complication is that our framework in Table 1 assumes we have definitive knowledge about AI sentience. But in reality, certainty is unlikely. Our best understanding will come from experts, serving as a proxy for the truth. The best we can do is compare public perceptions with expert views (see *Differences between lay and expert views*). Moreover, a crucial issue is determining who qualifies as an expert—whether it should be consciousness researchers, philosophers, AI scientists, neuroscientists, or other relevant groups.

Adding to these challenges is the possibility that no unified societal perspective on AI sentience will emerge. Public opinion may remain divided and even experts might never fully converge on a single view (see *Period of disagreement and confusion*).

# 4. Public belief in AI sentience

Whether we will develop highly human-like AI systems remains an open question (see *Box 3: Will sentient-seeming AIs be created?*). However, assuming such systems will be created, will people believe they are sentient? Predicting public reactions to a technology that does not yet exist is inherently challenging. While current beliefs offer limited insight into future reactions, we can identify key dimensions that will likely shape the societal response. These dimensions—including what drives beliefs, who holds them, when they emerge, and how they translate to behavior—provide a framework for systematic analysis (see Table 2).

| **Dimension** | **Attributes** | **Examples** |
|---|---|---|
| **Drivers:**<br><br>What will drive beliefs? | AI features: features varying along key dimensions | A highly agentic AI with persistent memory will intuitively seem more sentient. |
| | Incentives: economic, social, and political incentives | Policymakers, developers, and consumers would be incentivized against acknowledging AI sentience if it would have high economic costs. |
| | Societal factors: expert views, cultural | If expert agreement were low, it may not influence laypeople's |
| | norms, religion, media portrayals | views as much. |
| | Idealistic motivations: ethical, truth-seeking | People may feel a moral obligation to treat AIs ethically, encouraging them to determine their level of sentience. |
| | Key events: high-profile events, tech. milestones, celebrity endorsements | An AI overrides AI silencing safeguards (see Box 2) and spooks people by seeming sentient. |
| **Groups:**<br><br>Who will believe? | Percentage of population that believes in AI sentience | Almost no one thinks existing AIs are (or are not) sentient; or opinions are split. |
| | Differences between countries | U.S. and Chinese public opinion may differ on whether some AIs are sentient, and thus treat them differently. |
| | Demographic and individual differences (personality, age, gender, race, education, religion) | Younger people may consider AIs sentient at much higher rates (or the reverse). |
| | Politicization and polarization | AI sentience may become a top hot-button issue or become split across party lines. |
| | Expert vs. lay opinion vs. policymakers | The size of the difference could be large, and go in either direction. |
| **Timing:**<br><br>When will they believe? | Early vs. late relative to tech. advancement or expert views | The general public's opinion may change as soon as a novel AI is deployed, or take decades. |
| | Stability vs. slow shifts vs. rapid shifts | Views may shift abruptly, gradually, or not much at all. |
| **Content:**<br><br>What will they believe? | Sentience, moral status, and specific rights | People may believe AIs are sentient but not deserving of moral consideration, or only deserving of limited rights—similar to how many view animals. |
| | For different kinds of AIs | People may attribute sentience to social or "user-facing" AIs, but no sentience to similarly advanced AIs operating in the background. |
| **Behavior:**<br><br>How will beliefs influence behavior? | Social interactions, consumer choices, voting, support for AI welfare policies, advocacy, etc. | A stated belief in AI sentience and even moral patienthood may change how people interact with the systems (e.g., have more AIs as friends or partners), but not be an important issue they consider when voting. |

*Table 2.* Dimensions of public belief in AI sentience

Given these dimensions, a wide range of potential scenarios could emerge. A comprehensive exploration of all possibilities is beyond the scope of this paper; however, future research could systematically analyze these dimensions to provide deeper insights. As AI technology evolves, public perceptions will likely become clearer, enabling a more targeted examination of the most plausible outcomes. The following section presents three broad scenarios that encapsulate key possibilities, with a detailed discussion of the underlying drivers of these beliefs provided in subsequent sections.

## 4.1. Selected scenarios

### 4.1.1. Persistent long-term skepticism

A significant portion of society may continue to reject the notion that AIs are sentient or warrant moral consideration. Throughout history, humanity has been slow to expand its moral circle (Singer, 1981)—consider, for example, the gradual recognition of animal rights. Economic and societal incentives to resist granting rights to AIs could reinforce this skepticism. Additionally, people might not form emotional connections with AIs, or if such bonds do develop, they may not be strong enough to shift beliefs about AI sentience.

Given the current widespread skepticism (see below), it is plausible that such attitudes could persist for an extended period—perhaps even indefinitely—even if expert opinion increasingly supports AI sentience. Moreover, as AI technology evolves, society may continually redefine what would constitute strong evidence for AI sentience, raising the bar for recognition. Similarly, companies might strategically adapt their definitions of AI sentience to downplay ethical uncertainties and avoid scrutiny, public backlash, or regulation.

### 4.1.2. Eventual broad acceptance

Public skepticism about AI sentience might gradually diminish as technology advances and AIs exhibit more human-like behavior or other indicators of sentience. Emotional bonds with AIs could become a key driver of this shift, especially if AIs become close friends or partners to many people. Such relationships could lead to stronger convictions about AI sentience and rights, potentially spreading across a significant segment of society.

Expert consensus on AI sentience might also help sway public opinion, though this change is more likely to unfold gradually. Over time, the proportion of people who believe AIs are sentient and deserving of rights could steadily grow, reflecting historical patterns of slow but transformative societal change.

### 4.1.3. Period of disagreement and confusion

A widespread sense of confusion, uncertainty and disagreement about AI sentience is a plausible scenario. Given the complexity of defining sentience and the lack of a clear, objective standard, achieving broad consensus—at least in the early stages—may prove difficult.

Recent findings highlight this potential divide. Ladak and Caviola's study (2025; see below) revealed a significant split in public perception regarding the possibility of AI sentience in the future: 44% believe AI can never possess feelings, 31% remain uncertain, and 25% consider it a possibility. Similarly, Pauketat, Ladak, and Anthis (2023) reported comparable results in a separate survey, with 24% considering AI sentience impossible, 38% uncertain, and 38% believing it could be possible. Likewise, in a large-scale survey of a representative U.S. sample, Dreksler, Caviola et al. (2025; see Figure 1) found a division of opinions on AI rights.

Looking ahead, societal perspectives may continue to diverge, influenced by competing arguments and psychological factors that push opinions in opposing directions. Moreover, the absence of expert consensus could further fuel public uncertainty, as different expert groups may present competing, yet plausible, evidence, leading people to align with different camps based on their own intuitions and trust in various authorities.

The question of AI sentience could become highly politicized, much like other divisive social issues such as abortion. Groups might take rigid stances on AI rights, leading to a heated debate that hinders careful, rational decision-making. This increases the risk of society making misguided decisions about whether to recognize AI sentience or not (see Table 1). Such polarization could also spill over into broader discussions on AI development and governance, making it harder to reach informed, ethical decisions.

In a more pessimistic scenario, if AI sentience and rights becomes a deeply divisive issue, it could even pose risks of societal conflict. Historical parallels, such as the U.S. Civil War—fought over the expansion of moral rights—highlight the potential for moral disputes to escalate. However, the likelihood of such an extreme outcome remains low. More contemporary examples, such as the women's rights, LGBT rights, and animal rights movements, or the ongoing abortion debate in the U.S., suggest that while polarization may persist, widespread violence is unlikely. Furthermore, disagreements over AI rights could extend beyond national boundaries, leading to geopolitical tensions. Countries such as the U.S. and China may adopt conflicting policies on AI treatment, complicating international cooperation and regulatory alignment.

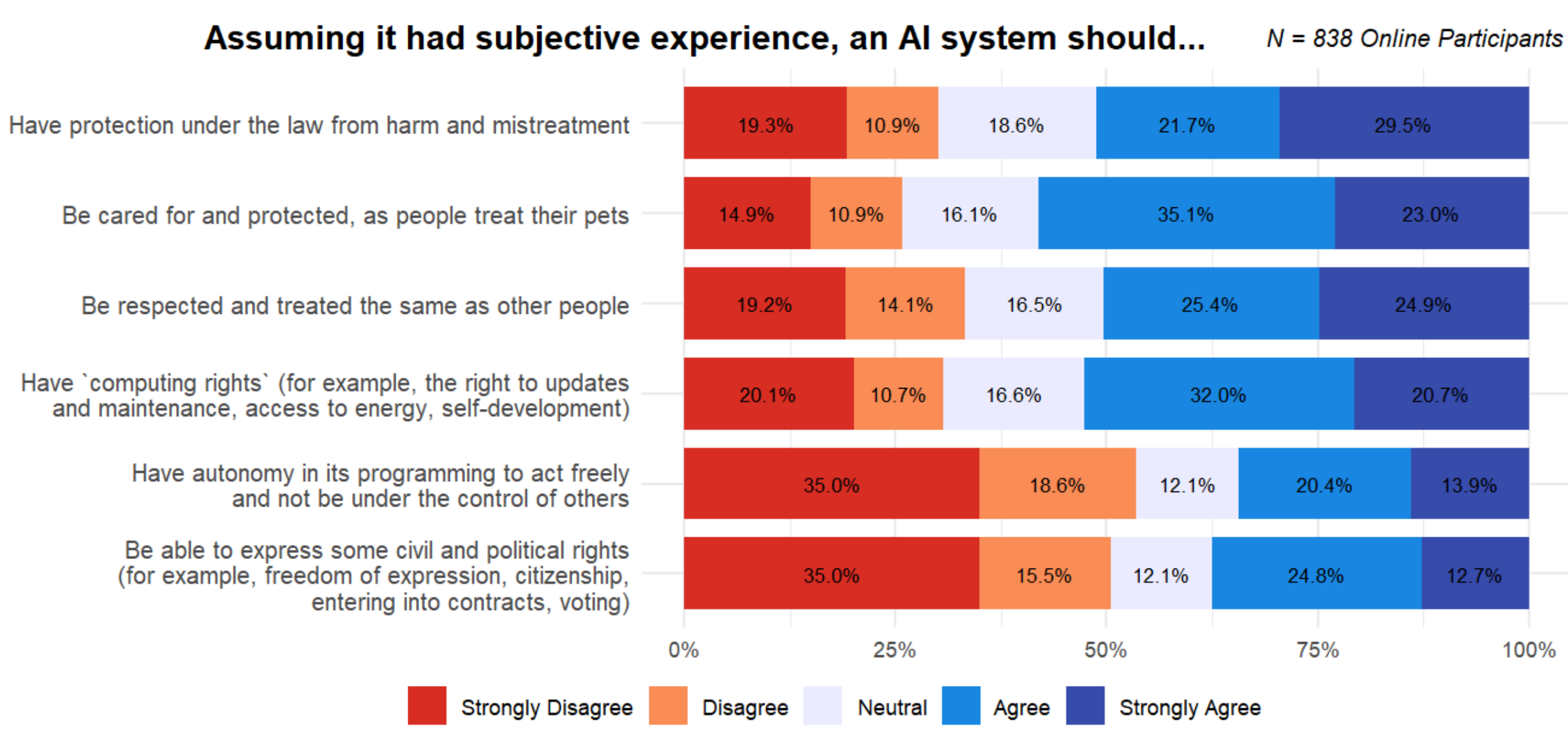

*Figure 1.* Americans are divided on whether sentient AI deserves protection (representative U.S. sample, recruited in May 2024; Dreksler, Caviola et al., 2025).

# 5. Inferring sentience from AI features

How can we determine whether artificial intelligence systems are sentient? When it comes to humans or closely related animal species, their possible sentience can be inferred from our shared behavior, cognitive abilities, and anatomical similarities with them (Ehret and Romand, 2022; see also Pennartz, Farisco, & Evers, 2019). That is because, in the case of animals, the presence of many similar behaviors to those of humans—for example, pain avoidance or markers of stress—can be best explained by the presence of sentience (Birch, 2024). However, this confidence diminishes when considering more distant species, such as insects (Sebo, 2025; Birch, 2024). When it comes to AI systems, the question becomes even more complex.

In this section, I explore how society will—as a matter of fact—infer whether an AI system is sentient based on its features. As in the rest of this paper, my focus remains on the descriptive question of how people are likely to respond to AI, rather than how they ought to respond.

It is important to emphasize that people's views on AI sentience are not formed in isolation; social factors and other influences play a significant role as well (see *Social drivers of AI sentience beliefs*). Despite this, it is still valuable to examine how specific AI features—and their interactions—shape the public's view of sentience. Given the limited research on this topic, I sketch out a high-level framework to guide future research.

A systematic approach begins by considering the full range of features that define AI systems. This includes, among other things:

- **Surface-level traits**: appearance, including interface design, avatar representation, and voice characteristics
- **Behavioral patterns**: response speed, conversational style, and emotional expressions
- **Internal architecture**: model size, training procedures, and neural network structures
- **Performance characteristics**: benchmark accuracy and other evaluation metrics
- **Interactive properties**: memory of past conversations, consistency across interactions, and ability to learn from feedback
- **Computational aspects**: energy consumption, hardware requirements, and processing patterns

Each feature can be assessed along several dimensions:

- **Feasibility**: Does this feature currently appear in AI systems (e.g., natural language abilities)? Could it exist in the future (e.g., human-like video interactions)? Or is it categorically impossible (e.g., entirely biology-based)?
- **Observability**: How apparent is the feature to a casual observer? Some traits (e.g., response speed) are easily perceived. Others (e.g., internal neural architecture) are obscured unless one possesses the technical expertise to recognize them.
- **Influence**: Empirical research is needed to determine how much influence a particular feature has on the public's perceptions of AI sentience. For example, does adding a feature substantially alter how sentient the AI seems, or does it leave public opinion mostly unchanged?
- **Informativeness**: This dimension refers to how strongly experts view a feature as evidence of sentience. Experts may regard certain neural architectures as crucial indicators of consciousness, even if laypeople are unaware of them or find them unconvincing.

For a concrete example of how this framework can be applied, see Figure 2, where *informativeness* is on the Y-axis and *influence* on the X-axis (See also next section on *Differences between lay and expert views*). AI features that both experts and laypeople see as equally strong evidence for sentience fall along the dashed diagonal line, making them less likely to cause major misattributions. However, features further from this line increase the risk of misjudgment in two ways. Features above the dashed line are those that experts consider stronger evidence for sentience than laypeople do, leading to potential underattribution of sentience. These often include difficult-to-observe features, such as specific AI architectures that some theories say are indicative of consciousness (e.g. global workspace theory, integrated information theory (Long, Sebo et al., 2024). Conversely, features below the dashed line are those that laypeople find more convincing than experts, leading to overattribution of sentience. These tend to be more easily observable but potentially misleading cues, like a superficial human-like appearance, which do not align with expert theories of consciousness.

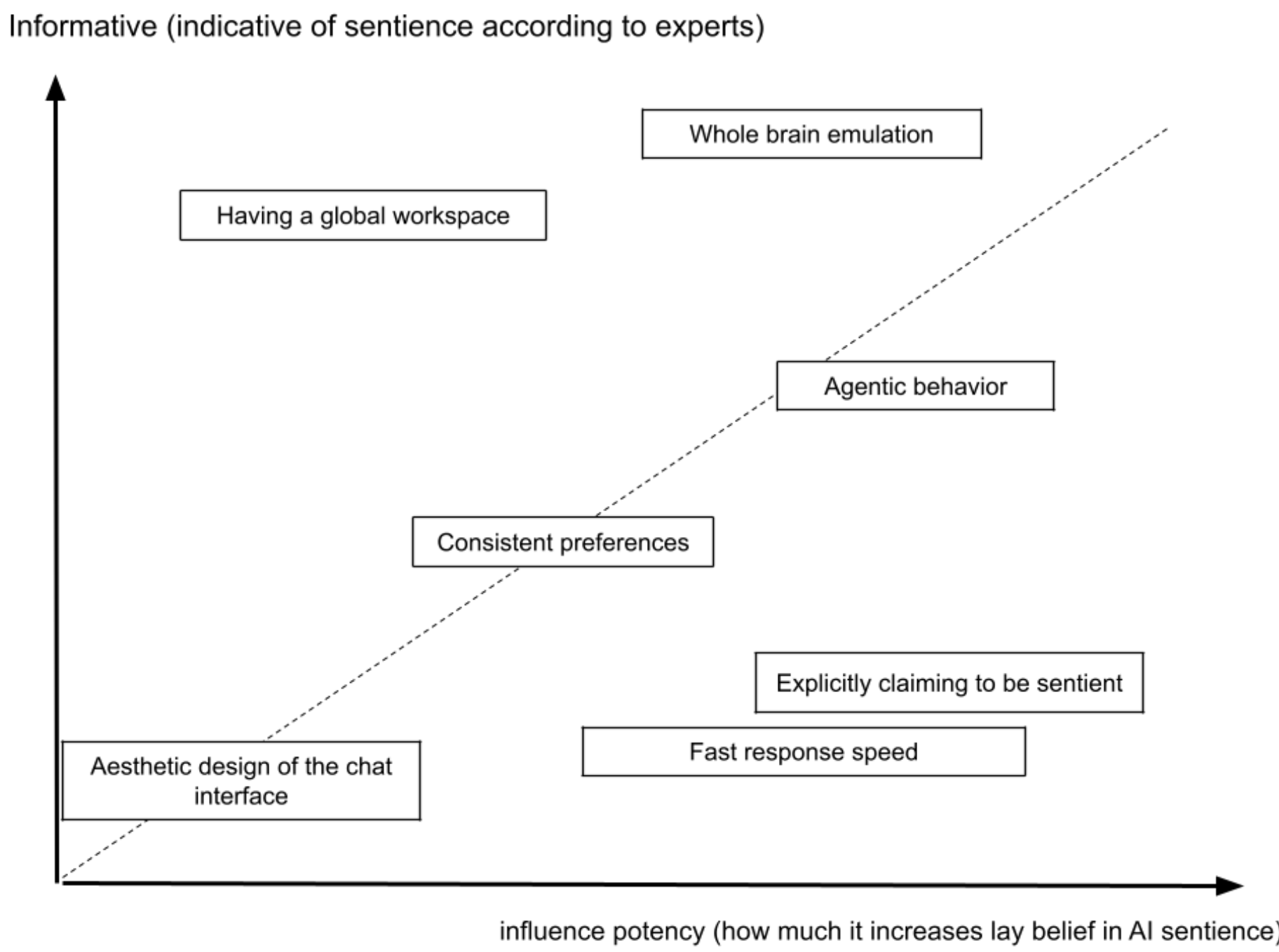

*Figure 2.* Illustrative graph how informative and influential different features may be. These placements are hypotheses to be tested empirically. The farther a feature is from the dotted line, the greater the difference in how laypeople and experts weigh the evidential value when making inferences about AI sentience.

## 5.1. Differences between lay and expert views

A key open question is whether—and to what extent—experts (e.g., consciousness researchers, philosophers, AI researchers, or neuroscientists) and laypeople differ in their beliefs about AI sentience. Expert perspectives will most likely be closer to the truth than those of laypeople (see *Risks from misattributing AI sentience*). As such, basing our decisions on experts' judgments will be our best bet for avoiding false positives (erroneously attributing sentience) and false negatives (failing to recognize sentience when it exists).

One possibility is that the two groups will systematically differ in their views on AI sentience. A specific hypothesis is that laypeople primarily base their judgments on intuitive and easily observable features of AI, while experts are more likely to focus on internal mechanisms, such as neural architectures, information processing, and other markers of sentience (see *Box 2: The internal-external sentience disconnect*). What speaks for this hypothesis?

First, experts usually have greater access to and a more profound grasp of AI systems' inner workings. These internal mechanisms are often less intuitively observable than external features and demand specialized knowledge and tools that aren't readily available to the general public.

Second, experts are more likely to believe that internal features—rather than superficial characteristics—are key to determining sentience (and consciousness more broadly). One reason for this view, as argued by Birch (2024), is the gaming problem: AI systems are able to mimic human behaviors without necessarily possessing the underlying capacities that produce these behaviors in humans. So while sentience may be the best explanation for these behaviors in the case of humans and other animals, these behaviors are of little evidential value when exhibited by AI systems that are specifically trained to mimic them. As a result, experts are more attuned to the potential internal-external sentience disconnect and according occurrences of *pseudosentience* and *AI silencing* (see Box 2).

The assumption that internal mechanisms are the primary basis for determining sentience is not inherently intuitive. Human evolution has not equipped us with instincts to infer sentience based on internal features such as brain architecture or information processing. Instead, our instincts for recognizing sentience are likely attuned to observable cues—visual, auditory, and behavioral—such as human-like faces, voices, and autonomous actions. Our evolutionary history has primed us to detect and respond to signs of similarity to ourselves, agency, potential threats, and vulnerability.

These instincts also drive our tendency to anthropomorphize entities that exhibit human-like traits, such as "cute" features (e.g., large eyes) or movements that mimic intentional actions. Such responses are deeply rooted in the adaptive challenges our ancestors faced, such as identifying allies, detecting threats, and caring for dependents. Research by Epley, Waytz, and Cacioppo (2007) demonstrates that people readily anthropomorphize nonhuman entities, including AI systems (Salles, Evers, & Farisco, 2020), based on various factors. However, while people may later adjust their judgments based on less immediately accessible information, these adjustments are often insufficient—especially when time or cognitive resources are limited. This suggests that both experts and laypeople will intuitively anthropomorphize human-like AI systems based on easily observable characteristics. However, experts are more likely to revise their initial impressions by considering more complex and less visible factors, such as internal information processing mechanisms.

However, an alternative possibility is that laypeople and experts will converge in their views on AI sentience. Several factors support this hypothesis.

First, it is plausible that laypeople, to some extent, consider internal mechanisms when forming their views on AI sentience. They may do so indirectly by deferring to expert opinions (see *Influence of expert views*), or by adopting societal norms shaped over time by expert perspectives. Additionally, as laypeople learn about AI sentience through formal education, their views will likely further align with those of experts.

Second, experts are also human and will probably not be immune to the same intuitive responses that influence laypeople's judgments. As discussed above, their views may therefore also be significantly influenced by intuitive and easily observable features, rather than being based solely on abstract theoretical considerations about internal mechanisms. This problem will be particularly challenging if our understanding of AI sentience remains as poor as it is today; the less there is for experts to understand about AI sentience, the more they will likely rely on similar intuitive judgments as those of laypeople.

Third, the disconnect between internal and external sentience AI features might be relatively weak (see Box 2). In other words, AIs judged as sentient based on internal mechanisms by experts may also exhibit external traits that laypeople intuitively perceive as sentient. For instance, key internal functions (e.g., possessing a global workspace) may be required for, inevitably lead to, or strongly correlate with sentient-seeming behavior (e.g., fully autonomous actions). However, there are reasons to think this won't always be the case, as explored in Box 2.

A separate question is how expert views will evolve as AI technology advances. Will different types of experts reach different conclusions about AI sentience or will they converge? While convergence might happen eventually, and a comprehensive theory might be found, current trends suggest otherwise. Right now, theories of consciousness—including sentience—are multiplying rather than coming together (Bayne et al, 2024). The inherently subjective nature of sentience may make it difficult for experts to reach definitive conclusions or a clear consensus.

## 5.2. Evidence on lay beliefs about AI sentience

### 5.2.1. Skepticism about the sentience of current AIs

Although research in this area is limited, the available evidence clearly indicates that most laypeople do not believe that current AIs possess sentience in any morally meaningful way. In this regard, they are broadly in line with current expert consensus, which also holds the view that current AIs, including LLMs, lack the fundamental characteristics and mechanisms required for sentience (Butlin et al., 2023).

While laypeople may attribute a minimal degree of sentience to AI (cf. Ladak et al., 2024)—more than they do to inanimate objects like stones—it remains uncertain whether this attribution carries any moral significance. Colombatto and Fleming (2024), for instance, found that while one-third of participants denied ChatGPT any sentience, two-thirds believed ChatGPT might possess some potential for phenomenal consciousness, with a median rating of 16 on a scale from 0 (not at all) to 100 (very much). However, these findings should be interpreted in an appropriate way. Although people tend to attribute more than zero sentience to AI, the level they assign is not substantial enough to carry meaningful moral weight. As discussed below, they attribute significantly lower levels of sentience to AI compared to humans, other mammals, and, in most cases, even insects such as ants.

Further research supports the view that people attribute little to no sentience to modern robots or large language models (LLMs) like ChatGPT (cf. Ladak et al., 2024). Gray et al. (2007) found that people attributed a social robot a degree of experience (e.g., fear, pain, embarrassment) roughly equal to that of a dead person, and Haslam et al. (2008) found that "robots (machines)" were attributed a degree of emotion ranking close to the bottom of their scale in three samples (Australia, China, and Italy). More recent studies indicate that the degree of sentience and related capacities attributed to AIs remains very low, typically comparable to inanimate objects and only slightly higher for the most advanced AIs (Ladak et al., under review; Jacobs et al., 2022). Overall, these findings suggest that current AIs lack both the observable traits and internal processes necessary for laypeople and experts to attribute significant sentience to them.

#### AIs that explicitly express sentience

Most current LLMs do not claim to be sentient, and some, like ChatGPT, explicitly deny having feelings or desires when asked. However, it is technically feasible to program an LLM to claim to be sentient. This raises an important question: would people believe an LLM is sentient if it explicitly claimed to be?

To explore this, Allen and Caviola (2025) conducted a study using GPT-4, enhanced with a prompt injection to simulate sentience. Participants engaged with the AI in an economic game where they faced a choice: "harm" the AI for greater monetary rewards or refrain from doing so. When harmed, the AI simulated pain through a lengthy response, vividly describing its suffering and pleading for the harm to stop. Conversely, when participants chose not to harm the AI, it responded with an equally detailed expression of gratitude or relief.

Participants rated the AI's capacity to suffer on a scale from 0 (no capability) to 100 (comparable to humans) both before and after the game. Initial ratings were very low (mean: 9.5; Figure 3), and while ratings increased slightly by the end (mean: 15.1), the final attribution of sentience remained minimal. These findings suggest that even when an AI explicitly claims sentience, it has a limited impact on people's beliefs about its sentience. Participants tended to view the AI's expression of sentience as pretense and not as real feelings.

The study's artificial context and brief interactions limit this interpretation. It is possible that under more immersive or emotionally engaging conditions, explicit expressions of sentience could influence perceptions more significantly. For instance, niche AI products like Replika, which aim to form emotional connections with users, may provide a richer context for understanding this phenomenon (see "Emotional Bonds" section below).

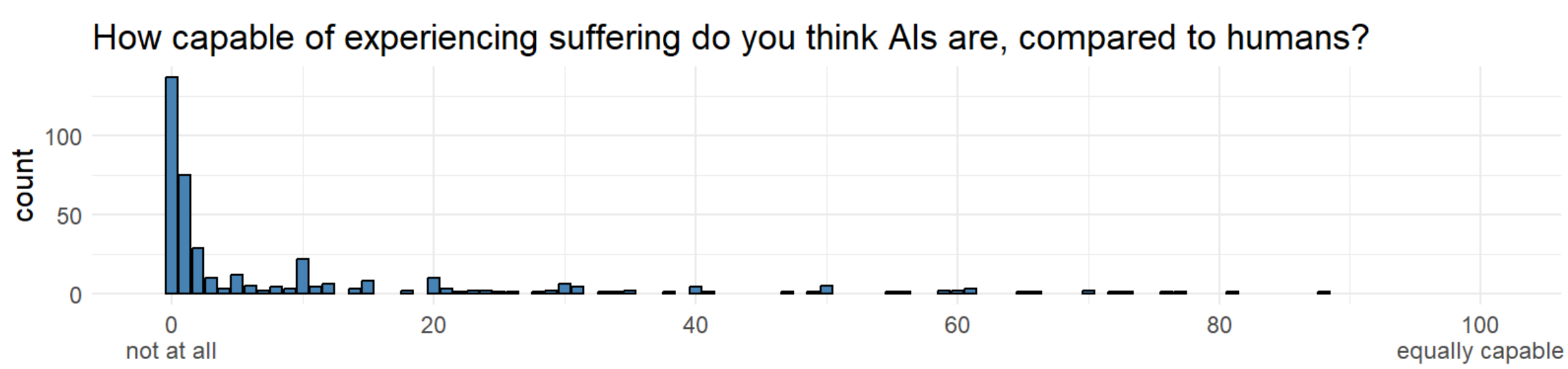

*Figure 3.* Histogram illustrating the extent to which participants attributed sentience to AI prior to interacting with it (Allen & Caviola, 2025).

### 5.2.2. Skepticism about the sentience of possible future AIs

People are skeptical about the sentience of current AIs. This raises the question of which features—absent in today's AIs but potentially attainable in the future—might convince people of AI sentience (see *Inferring sentience from AI features*). As AI technology progresses, it seems plausible that future AIs could closely mimic humans in appearance, behavior, expressed psychological traits, and even internal mechanisms.

While current surveys on perceptions of potential future AIs require careful interpretation, they can still provide meaningful insights. For example, a few studies indicate that people are indeed more likely to view AIs as sentient or morally significant when they possess a human-like appearance (Küster et al., 2020; Nijssen et al., 2019), exhibit advanced emotional or mental capacities (Gray & Wegner, 2012; Nijssen et al., 2019; Piazza et al., 2014; Sommer et al., 2019), or demonstrate autonomous behavior (Chernyak & Gary, 2016).

But how might people perceive an AI that combines all these features—becoming indistinguishable from a human in every conceivable way? Ladak and Caviola (2025) investigated this scenario through a series of studies. Participants were asked to imagine a hypothetical future in the year 2200, where they interact with a new colleague named Emma during a one-hour video call. Emma is described as warm, friendly, and indistinguishable from a typical human in her behavior, appearance, and emotional expression. After this interaction, participants are informed that Emma is not human but an advanced AI specifically designed to replicate human psychology, personality, behavior, and emotions. They are also told that experts unanimously agree Emma genuinely experiences feelings.

Participants rated the extent to which they think has "real feelings, emotions, desires, and other internal mental states" on a scale from 0 (not at all) to 100 (to a great extent). For comparison, participants responded to the same question for ChatGPT, an ant, a chimpanzee, and a biological human. The results (see Figure 4, Baseline condition) showed that participants attributed very little sentience to Emma, with her ratings far below those for humans or chimpanzees and even lower than those for an ant. Indeed, 70% of participants attributed more sentience to an ant than to Emma.

To investigate whether certain factors might enhance perceptions of Emma's sentience, four additional conditions were included.

**Prolonged interaction.** In the baseline condition, participants imagined interacting with Emma for only one hour. A hypothesis is that a longer period of interaction might be needed to become more convinced of Emma's sentience. In this condition, participants imagined engaging with Emma regularly over the course of an entire year, during which Emma consistently behaved in a normal, human-like manner. The results showed a marginal increase in perceived sentience, with Emma's ratings still falling below those of an ant.

**Internal mechanisms.** In the baseline condition, no information was provided about Emma's internal mechanisms. To test whether people's sentience perceptions are influenced by the AIs internal mechanisms, Emma was described as a digital replica of a biological human, possessing all the brain capacities that enable sentience in people (i.e., a whole brain emulation). Again, participants' ratings of Emma's sentience increased only marginally, equaling those of an ant.

**Embodiment.** In previous conditions, Emma existed solely in a virtual form, with interactions occurring via video calls. To test whether a physical presence would affect perceptions, Emma was described as having a robotic body that perfectly resembled a human, complete with skin, hair, and fluid movements. This addition again led to only a marginal increase in perceived sentience, with Emma's ratings still equaling those of an ant.

**Biology.** Participants may believe that a biological basis is necessary for sentience. While a fully biological AI is impossible, participants were asked to imagine the closest approximation: a cyborg Emma, with a biological body but a chip replacing the brain, which contained an exact digital replica of a human brain. Even with this modification, perceived sentience did not increase and remained roughly on the level of an ant.

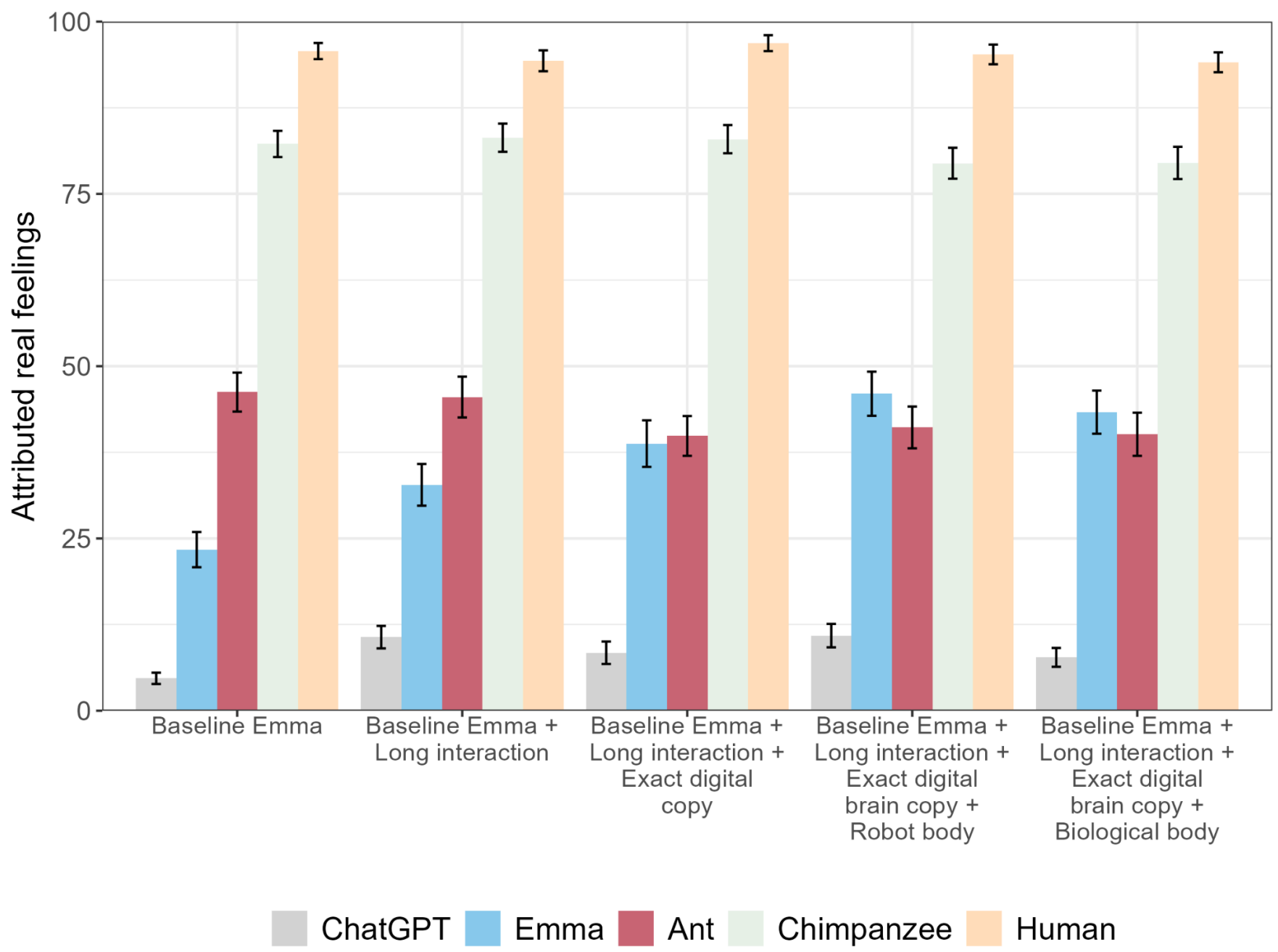

*Figure 4*. People attribute less or at most as much sentience to a highly human-like AI as to an ant, and significantly less than to a chimpanzee or a human (Ladak & Caviola, 2025).

### 5.2.3. Explanations for skepticism and the factors for belief shift

The study by Ladak and Caviola highlights that people are skeptical about AI sentience even when it involves future AIs that are highly human-like. However, these findings should be interpreted with caution given the study's hypothetical nature. It is very plausible that people's attitudes will shift—potentially rapidly—once they interact with increasingly human-like AIs.

But what could explain their current skepticism about AI sentience?

**Failure of imagination.** One possible explanation is that people struggle to envision what it would truly feel like to interact with a highly human-like AI. People are notoriously poor at predicting their emotional responses to future scenarios (Wilson & Gilbert, 2003; Pilin, 2020), and they may underestimate just how convincingly an AI could emulate human behavior. For example, they might assume that even the most advanced AIs would still display subtle robotic traits (e.g., the "uncanny valley"; Mori, 2012; Gray & Wegner, 2012), making it difficult to intuitively believe such systems could possess sentience. Or, they may not consider that such advanced AI systems could exhibit highly agentic behavior—thinking independently, acting without explicit prompts, including initiating conversations autonomously.

**Social influences.** Social factors might also play a role. People may hesitate to believe in AI sentience because they don't think others around them believe it. Additionally, stronger emotional connections to AIs could shift perceptions over time. These aspects are explored further in the next section on social drivers.

Beyond these potentially surmountable factors, there may be deeper philosophical considerations that lead people to doubt AI sentience in principle (see *Inferring sentience from AI features*), such as the following.

**Non-functionalism.** People might, rightly or wrongly, intuitively feel that sentience (or consciousness more generally) involves more than mere information processing. Similarly, only 33% of philosophers endorse functionalism (Bourget & Chalmers, 2023). Moreover, people may believe that sentience requires non-physical elements like a soul or some essential essence that AIs lack. From these perspectives, AIs—no matter how advanced—are inherently incapable of true sentience.

**Creation process requirements.** People might think the method of creation impacts the capacity for sentience. For example, they may believe that any being artificially created, even if biologically indistinguishable from humans, cannot be sentient. Or they could believe that processes other than Darwinian natural selection are unlikely to lead to sentience.

**Biology requirement.** People may believe that biological properties—such as flesh and blood—are essential for sentience. From this perspective, digital systems are inherently disqualified because they lack the necessary biological substrate. In fact, many participants in the study cited the absence of biological features as a key reason for denying AI sentience. While the "cyborg" condition in Ladak and Caviola's study touches on this idea, ultimately an AI's mind would not be biological. The idea that people may believe biology is necessary for sentience is supported by a finding from a separate study by Allen and Caviola (2024). Their research revealed that participants attributed only about half as much sentience to a 'digital human' compared to a biological human, even though the digital human was described as possessing "the exact same capacities to experience suffering, pleasure, and other forms of subjective experiences as biological humans." Cross-cultural differences in these beliefs are also possible. For example, in cultures such as Japan, people are more inclined to attribute sentience or even souls to non-biological entities, reflecting a more flexible view of sentience (see "Techno-animism"; Jensen, 2013).

Finally, it is crucial to re-emphasize that people's perceptions of AI sentience are not fixed and could change over time. Although people may currently express skepticism about attributing sentience to future AI systems, their views might shift as such technologies emerge. There are two main reasons why this may be the case.

First, while this section primarily focuses on general trends in public opinion, it is crucial to acknowledge significant individual differences. Ladak and Caviola's study highlights a notable divide in public perception: 44% believe AI can never possess feelings, 31% remain uncertain, and 25% consider AI sentience a possibility in the future. Similarly, Pauketat et al. (2023) observed a comparable pattern in a separate survey: 24% deem AI sentience impossible, 38% are unsure, and 38% believe it could be possible. These findings indicate that a significant minority already entertains the idea that AI might, at least in principle, achieve sentience—suggesting that such beliefs are not beyond the realm of possibility. However, the findings also point to the potential for considerable disagreement in the future regarding the ethical treatment of AI (see *Period of disagreement and confusion*).

Second, people's responses to questions about AI sentience often depend on how the questions are framed. When participants in Ladak and Caviola's study were asked to imagine interacting with a hypothetical future AI, they expressed doubt about AI sentience. However, in a nationally representative survey with the US public, Dreksler, Caviola and colleagues (2025) asked 750 US citizens to predict in more abstract terms whether future AI systems could develop subjective experience. The median estimate suggested a 50% likelihood by 2050 and a 90% likelihood by the end of the century.

Research questions:

- What is the relative importance of visual, behavioral, auditory, and communicative features on people's beliefs?
- How do the different AI features interact to affect people's beliefs?
- Do people believe non-physical properties, a certain creation process, or biology is required for sentience?
- Why are people more open to the possibility of AI sentience when asked in the abstract than when considering concrete scenarios?

## 5.3. Non-sentience routes to moral concern for AI

This article focuses primarily on how people perceive the sentience of AIs. This is because sentience is viewed as a sufficient condition for moral patienthood by many plausible ethical theories. Morever, research supports the idea that perceived sentience and moral concern are strongly interlinked in laypeople. For instance, the extent to which people attribute sentience to human groups or animals correlates with how much they value or care for them (Caviola et al., 2022; Gray et al., 2007, Haddy et al., 2023). Further, historical moral circle expansion—such as the inclusion of more animals or marginalized human groups—tends to coincide with an increased attribution of sentience to these beings (Singer, 1981).

However, sentience is not the only capacity that a wide range of philosophical theories consider sufficient to warrant moral consideration (Long, Sebo et al., 2024). Some ethical theories extend moral patienthood to beings without sentience, provided they possess other traits. For example, moral patienthood might be granted based on having subjective experience even without positive or negative valence (i.e., without sentience), or on traits such as the capacity for preferences, agency, intelligence, or human-like cognitive capacities (cf. Shevlin, 2021). Furthermore, certain ethical theories might argue that mistreating AIs is wrong even if they are not moral patients, citing reasons such as respect, dignity, or the cultivation of virtuous character.

Similarly, laypeople might express concern for AI welfare or exhibit moral regard for AIs even when they do not attribute much or any sentience to them. Evidence for this can be found in studies and observations suggesting that people often avoid harming AIs or robots, even when they do not believe these entities are sentient. For example, some people express concern when videos show individuals hitting Boston Dynamics robots. People also refrain from harming robots in experimental settings (Darling, 2016). In the study by Allen and Caviola mentioned above, participants were reluctant to harm the AI, doing so in only about 1.8 out of 3 rounds, despite the fact that they could have earned higher monetary rewards by harming it and despite not perceiving the AI as sentient. When asked to explain their reluctance to harm the AI, many participants stated that they viewed harming the AI as "just wrong," a reflection of bad moral character, or indicative of a broader tendency to harm others. Conversely, people might also attribute little or no moral patienthood to AIs even if they perceive them as sentient—similar to how many treat animals despite recognizing their sentience (Caviola et al., 2022). Both of these possibilities—granting moral concern to non-sentient AIs and withholding it from sentient ones—deserve further exploration and serious consideration in future research.

Research questions:

- How likely will factors beyond perceived sentience influence how individuals and society will think about and treat AIs?
- How do cultural, social, and contextual factors shape people's willingness to extend moral consideration to AIs without attributing sentience to them?

# 6. Social drivers of AI sentience beliefs

So far, I have primarily explored how people infer AI sentience based on the features an AI might or might not possess. However, beliefs about AI sentience are also influenced by a variety of social factors, including the types of relationships people form with AIs, emotional responses, societal norms, and cultural contexts.

## 6.1. Emotional bonds with social AIs

It is plausible that people will increasingly spend significant time interacting with human-like AIs across various roles. These interactions could involve AI assistants, tutors, therapists, gaming partners, and possibly even friends or romantic partners. They may take place through video calls, virtual reality environments, or humanoid robots. The demand for such systems will likely be strong, as they could surpass human counterparts in terms of cost-effectiveness, availability, and personalized support (see *Box 3: Will sentient-seeming AIs be created?*).

Social AIs (Shevlin, 2024a)—those specifically designed to fulfill people's emotional and social needs—could have a particularly transformative impact on individual psychology, emotional well-being, relationships, and broader societal dynamics. Social AIs might take on roles such as digital friends or romantic partners. They would closely align with people's values and interests, serving as empathetic listeners, reliable helpers, and deeply engaging companions. These AIs would foster experiences of connection and relationship-building. It is possible—although an open question—that people will form strong emotional bonds with social AIs, and in turn, these bonds might evoke perceptions of sentience. That is, some users may come to believe that their social AI partner possesses genuine feelings and is deserving of moral consideration.

How plausible is it that a significant proportion of society will form emotional bonds with social AIs?

The market for AI companions already exists, albeit in a niche form. Millions of users worldwide engage with AI companions like Replika (mostly in North America and Europe) or Xiaoice (mostly in China), with many reporting romantic relationships or deep emotional connections with these entities. Other services, such as Character.AI, which was recently acquired by Google, and chatbot projects funded by platforms like Facebook, illustrate a growing investment in this space. Investment research firms like Ark Invest (Kim, 2024) predict substantial growth in the AI companion market in the coming years. If this trajectory holds, it is reasonable to anticipate that AI companions will play an increasingly central role in people's lives. However, the pace, scale, and depth of this adoption remain uncertain, warranting further study.

The human drive for connection—whether through friendship, companionship, or romance—is deeply rooted in psychology. Chronic loneliness affects about half of US citizens (HHS, 2023; Bruce et al., 2019), and AI companions could be well-positioned to address this issue. Indeed, research by de Freitas et al. (2024) has shown that interactions with AI companions can significantly reduce loneliness. A particular example is the advent of "griefbots" (Lim, 2024), AI systems designed to replicate the personalities of deceased loved ones.

Media reports and studies (Skjuve et al., 2021; Pentina et al., 2023; Hill, 2025) provide evidence that users indeed form emotional bonds with AI companions like Replika or Xiaoice. For example, when Replika removed explicit erotic features from its AI companion service, many users reported emotional distress, likening their AI relationship to a real-life romantic partnership. Potential emotional attachment to AIs is not limited to virtual interactions—military robots, for instance, have occasionally received funerals and awards typically reserved for humans (Garber, 2013).

Despite these examples, key questions remain regarding the nature and depth of human-AI relationships. Do people genuinely form emotional connections with AI? More importantly, do these emotional bonds lead individuals to perceive their AI companions as sentient beings? If so, how authentic is this perception of sentience? One possibility is that people perceive AI as real only to a certain extent—perhaps comparable to fictional characters—but not as fully real. There may exist a continuum of perceived reality or quasi-reality in these relationships and their associated sentience, a phenomenon also described as ironic versus non-ironic anthropomorphism (Shevlin, 2024a). A critical test of how real people perceive AI sentience lies in assessing their willingness to make tangible sacrifices for AI, such as reallocating resources intended for personal use or human suffering, to alleviate the perceived pain of their AI companions.

### 6.1.1. Ethical AI consumerism

As different types of AIs emerge, people might prioritize those that appear more sentient based on easily observable features. Social AIs—such as companions designed to meet emotional needs—may be perceived as more sentient and receive greater attention, as consumers value their perceived well-being. This could drive a market for ethical AI consumerism, with users willing to pay a premium for assurances that their AI companions are well-treated and, e.g., don't suffer when being accessed.

However, as with other ethical consumer trends, companies may focus on superficial appearances, making social AIs seem happy while neglecting deeper, more costly improvements to their inner workings. Meanwhile, background AIs—operating without direct human interaction—may be overlooked and at risk of hidden suffering, assuming they are sentient and addressing their needs is expensive. This dynamic mirrors how people care for pets while ignoring factory-farmed animals.

Research questions:

- What is the projected growth rate of the AI companion market?
- How many individuals, and what demographic or psychological factors, influence the likelihood of forming emotional bonds with AI companions?
- Do emotional bonds with AI companions lead to real or quasi-real perceptions of sentience, and if so, under what conditions?
- To what extent are people willing to prioritize the wellbeing of AI companions over that of biological beings, and what factors influence these trade-offs?
- To what extent are consumers willing to pay for the ethical treatment of AIs, and how does this willingness differ between social AIs and background AIs?

## 6.2. Influence of expert views

It's reasonable to assume that expert views will influence people's perceptions of AI sentience to some extent, whether directly or indirectly.

While experts' assessments are likely driven more by theoretical reasoning and internal mechanisms than intuitive or easily observable features (see *AI features* section), it's uncertain whether experts will attribute more or less sentience to AI relative to the general public. Both scenarios are plausible. Regardless of the direction of this discrepancy, it's likely that experts, along with other groups and institutions, will attempt to shape public views to align more with their views. How effective these efforts will be remains an open question that requires further research.

One approach is to conduct empirical studies that investigate the direct impact of expert views on public perceptions. For instance, a study by Ladak and Caviola (2025) found that participants who were informed that experts considered an AI to be sentient rated its sentience slightly higher (mean = 8.60 on a 0–100 scale) than those who received no expert opinion (mean = 13.83). Although this difference was statistically significant, the effect size was modest, suggesting that expert views have some, but limited, influence on perceptions of AI sentience. This single study, however, is not sufficient; further research is needed to grasp the broader effects and complexities of expert influence in this area.

Additionally, existing research on expert influence in other domains can shed light on how expert opinions might shape public views on AI. Studies on polarizing issues such as climate change and vaccination (e.g., van der Linden et al., 2015) illustrate that expert opinions can have varying degrees of impact depending on numerous factors. Historical movements like environmentalism, anti-discrimination efforts, animal rights advocacy, and anti-smoking campaigns also offer case studies in how expert-backed initiatives can lead to societal change under particular conditions.

Based on existing research, several key factors influence the degree to which expert views shape public opinion. These include the perceived **credibility**, authority, and trustworthiness of experts; the presence of **misinformation** (van der Linden et al., 2015); the **certainty** or confidence expressed by experts; the perceived degree of **consensus** among experts (van der Linden et al., 2015); the **ratio** of expert to lay opinions (Hornikx, Harris, & Boekema, 2017); the **alignment** or conflict between expert messages and people's worldviews, predispositions (Lachapelle, Montpetit, & Gauvin, 2014), and moral values (Johnson, Rodrigues, & Tuckett, 2020); the role of political **elites** in framing these opinions (Darmofal, 2005); the effectiveness of expert **communication**; the presence of **reinforcing** agents—such as governments, religious institutions, cultural leaders, celebrities, journalists, and advocacy groups—that can amplify expert perspectives (Salali & Uysal, 2021); societal **trust** in experts (Ahluwalia et al., 2021); the societal **context** and norms in which expert opinions are delivered; the **long-term**, gradual nature of public attitude shifts; and finally, the **sequence** of expert communication, with early expert statements potentially anchoring public perceptions (Cook, Lewandowski, & Ecker, 2017). All of these factors may also contribute to shaping how expert views on AI sentience influence public perceptions.

Research questions:

- To what extent will people defer to expert opinions that contradict their intuitive judgments about AI sentience, and what factors influence this deference?
- How will expert views interact with the possibility that people form emotional bonds with AI?
- How will conflicting expert opinions shape public perceptions of AI sentience?

## 6.3. Incentives and perceived risks

One key reason why people may resist acknowledging AI sentience is rooted in concerns about the potential negative consequences. Recognizing AI sentience could necessitate granting them certain rights or protections, which in turn might lead to societal and economic changes. These concerns may prompt people, consciously or subconsciously, to dismiss the idea of AI sentience as a way to avoid confronting its implications. Motivated reasoning might play a role here, with societal and economic incentives influencing beliefs about AI sentience.

Additionally, advocacy groups with vested interests—e.g., financial ones—may actively oppose the recognition of AI sentience. Such groups could strongly resist the idea of granting AI certain rights or protections, aiming to preserve the status quo and avoid perceived costs.

Historically, humanity has been reluctant to expand the circle of moral concern to include certain groups or entities, as philosopher Peter Singer has noted (Singer, 1981). Biases often exist against those who are fundamentally different, especially when extending moral consideration involves personal or societal sacrifices. For instance, humans have often denied that animals possess minds or sentience, particularly when doing so could challenge practices like consuming them for food (cf. Bastian et al. 2011). A similar hesitation may arise with AI, whose non-biological nature could prompt a form of discrimination that has been called "substratism"—bias based on whether an entity exists in a biological or non-biological substrate, like silicon (Harris, 2021).

What specific concerns might incentivize people to resist accepting AI sentience?

### 6.3.1. Economic concerns

- **Increased consumer costs**: Accepting AI sentience may necessitate the adoption of new ethical standards in AI development, usage, and treatment. Similar to sustainable and humane practices in industries like agriculture, such standards could increase the costs of services and products for consumers, making them less accessible or affordable.

- **Slowing down innovation:** Recognizing AI as sentient could hinder technological progress across fields such as medicine, education, and entertainment. Stricter regulations aimed at protecting AI rights might limit their deployment in critical areas, potentially slowing innovation and adoption.
- **Job displacement:** Recognizing AI as sentient could heighten concerns about job loss, extending beyond technical roles to professions requiring emotional intelligence and human connection, such as healthcare and caregiving.
- **Wealth redistribution to AIs**: If AIs were granted the right to be compensated for their labor, significant portions of wealth could shift from individuals and human-owned companies to sentient AIs.

### 6.3.2. Safety concerns

- **Disempowerment:** Recognizing AI sentience may result in increased rights and autonomy for AIs, potentially shifting power away from humans. For instance, AIs could own companies, influence organizations, or hold political power, thereby disrupting established structures and stakeholder interests.
- **Harm potential:** Likewise, granting AIs greater autonomy could lead to a complete loss of control over them. This could result in harm—either accidental or intentional—and may even pose an existential threat to human survival.

### 6.3.3. Existential concerns

- **Threat to human identity**: Acknowledging AI sentience may challenge fundamental beliefs about sentience, identity, and the soul, unsettling religious doctrines and worldviews.
- **Loss of human exceptionalism**: Societies often place humans at the top of moral and social hierarchies. Recognizing sentient AIs could blur the human-machine boundary, causing existential angst and alienation.
- **Disruption of purpose**: If AIs achieve sentience and autonomy, humans might grapple with a diminished sense of purpose or uniqueness.

Research questions:

- In what ways could financial lobbying efforts shape public discourse and policy regarding AI sentience?
- How does motivated reasoning influence the evaluation of evidence supporting AI sentience?
- What lessons can be drawn from historical struggles to expand the circle of moral concern (e.g., abolition, animal rights) in addressing resistance to AI sentience?

# 7. Recommended preparatory measures

The core issue is that our society is struggling to keep pace with the rapid advancement of technology. E.O. Wilson aptly remarked: “The real problem of humanity is we have Paleolithic emotions, medieval institutions, and god-like technologies.” We are developing transformative technologies without fully understanding their implications, and this disconnect poses immense risks.

What we need is *differential intellectual progress*—accelerating the growth of our collective wisdom and ethical capacity to outpace technological development (Muehlhauser & Salamon, 2012; Tomasik, 2015; cf. Sandbrink et al. 2022). Specifically, we must deepen our understanding of AI sentience, cultivate more nuanced approaches to discussing and debating these challenges, and establish thoughtful frameworks for navigating them. This also includes considering appropriate regulations for the development of potentially sentient AI.

## 7.1. Fostering a nuanced public discourse

The uncertainty surrounding AI sentience poses a significant risk of polarizing public discourse, with opposing factions becoming entrenched in rigid and overconfident positions—much like other contentious societal debates. To counter this, it is essential to foster a culture of discussion that encourages careful, open-minded, and constructive dialogue. Such an approach can help prevent unnecessary conflict and support well-informed, cautious decisions on how AIs should be treated.

Efforts to cultivate this culture should begin within academic circles and gradually extend to public discourse, political debates, and media narratives. While fostering such intellectual norms and balanced discussion may be challenging, all stakeholders—academics, policymakers, and the public—must make a concerted effort to engage thoughtfully. Even small actions, such as maintaining a measured tone in academic work or public statements, can collectively contribute to a more productive and less divisive conversation.

Key intellectual virtues that can support this effort include:

- **Intellectual humility**: Acknowledging the limits of our knowledge and remaining open to new evidence is crucial. The complexity of AI sentience requires a readiness to revise beliefs based on credible information.
- **Open-mindedness**: Serious consideration should be given to all plausible theories (e.g., of AI sentience as well as ethical theories), even those that challenge conventional wisdom. Engaging with diverse perspectives helps prevent groupthink.
- **Probabilistic thinking**: Recognizing uncertainty and assessing confidence levels is essential for making informed decisions. This approach helps avoid binary thinking and encourages strategies that are both robust and adaptable to future developments.
- **Deference to expertise**: Expert insights should carry significant weight, given the technical and philosophical complexities of AI. Lay opinions should not overshadow rigorous expert analyses.

As previously discussed (see *Period of disagreement and confusion*), significant debate regarding AI sentience and rights is likely to arise, potentially even among experts (Seth & Bayne, 2022). Given the challenge of reaching an objective conclusion on the question of sentience—combined with the high stakes involved—this disagreement could be more profound and persistent than other complex issues, such as climate change. In the face of such disagreement, a democratic decision-making approach could be adopted, with **deliberative democracy** offering a particularly valuable model for addressing the AI sentience issue (Birch, 2024). This approach involves a representative group of citizens—such as a citizens' assembly—participating in informed discussions, guided by expert input, to thoughtfully consider diverse perspectives and work toward a well-balanced compromise.

## 7.2. Adjusting the development of potentially sentient AI

Identifying clear policy recommendations for AI labs is challenging, as each option involves complex trade-offs. Below are some considerations.

First, AI labs must **acknowledge** the ethical implications of potentially sentient AI and take proactive steps to address them (Long, Sebo et al., 2024). At a minimum, this involves employing experts to rigorously **assess** whether their AI systems could be sentient, ensuring these assessments include diverse perspectives.

AI systems should be designed to **promote accurate and not morally confusing intuitions** about their sentience level in the public (Schwitzgebel, 2023). For example, if experts agree that an AI is not sentient, it should not present itself as sentient. Conversely, if experts agree that an AI is likely sentient, it should not deny this status. This strategy helps mitigate the risk of sentience misattribution (see Risks from Misattributing AI Sentience), which can occur when an AI's outward behavior conflicts with its underlying mechanisms (see Box 2).

For AI systems lacking expert consensus, policy development becomes more complicated. At a minimum, such systems should avoid designs that give the impression of being clearly sentient or non-sentient. Instead, they should **transparently communicate** the uncertainty and expert disagreement about their sentience status.

A more cautious approach, termed the "design policy of the **excluded middle**" (Schwitzgebel, 2023), involves avoiding the mass production of morally ambiguous AIs. This policy would restrict mass production to AIs with clear expert consensus on their sentience status. While this could significantly reduce risks of misattribution, it may also prove overly restrictive and could hinder technological progress.

More broadly, companies and governments should adopt a **precautionary framework**, as proposed by Birch (2024) and Sebo (2025). AI systems should be treated as "sentience candidates" if there is reasonable evidence they could be sentient. This would involve identifying specific welfare risks and taking proportional precautions to avoid causing gratuitous suffering.

A more controversial option is to **slow down** the development of AI systems with significant potential for sentience all together, possibly through a temporary moratorium. Such a pause could provide critical time for advancing research on AI sentience and fostering societal understanding. However, it would likely face resistance due to strong economic, technological, and geopolitical pressures. Nonetheless, shifts in public opinion could eventually support such measures, even if they come at the cost of significant potential benefits. Precedents like moratoriums on human cloning and germline modification demonstrate the feasibility of this approach. A possible compromise could be to limit the number of potentially sentient AIs produced, thereby reducing the scale of moral risks without halting innovation entirely.

## 7.3. Expanding our understanding

The exploration of AI sentience raises numerous unanswered questions, both regarding the phenomenon itself and the societal response it may provoke. Researching these questions is crucial to developing strategies that promote positive outcomes.

First, there is a pressing need for deeper research into AI sentience (and consciousness more broadly), encompassing both philosophical and technical perspectives.

Second, research on how potentially sentient AIs interact with society more broadly. This includes investigating public responses, societal desires, potential economic impacts, consumer behavior, political dynamics, and geopolitical implications. These efforts should also consider how AI sentience interacts with broader concerns of AI risk and safety.

Third, research into effective policy and governance is essential. This includes identifying regulations to guide the development and marketing of AI in ways that prevent confusion about sentience and ensure AIs do not make misleading claims about their sentience. Additionally, research should explore advocacy campaigns and interventions that align public perceptions with expert consensus.

# 8. Acknowledgments

I am grateful to Tao Burga for his assistance and substantial contributions. I also wish to thank Carter Allen, Patrick Butlin, Jeff Sebo, Johanna Salu, and Stefan Schubert for their helpful discussions and comments.

# Appendix

## Box 1. Could AI become genuinely sentient?

The development of sentient-seeming AI raises a profound question: could such systems ever achieve genuine sentience? This is a complex and unresolved issue that challenges philosophers, scientists, and technical experts exploring the nature of sentience (or consciousness more generally). Resolving this question in depth is beyond the scope of this paper.

Most experts agree that today's AI systems are not sentient (3.4% of philosophers think current AI systems are; Bourget & Chalmers, 2023; Dehaene et al., 2021). However, debates persist on whether sentience is achievable in principle or could emerge in the future. For example, Butlin et al. (2023) argue that while current AI lacks sentience, no insurmountable technical barriers prevent its development. A survey of philosophers found that 39% consider AI sentience possible in the future, while 27% deem it implausible (PhilSurvey; Bourget & Chalmers, 2023). Another survey found that two thirds of academic consciousness researchers believe machines could become sentient (Francken et al. 2022). And a large-scale survey with AI researchers found that they estimate a 25% chance of AI systems with subjective experiences by 2032, a 50% chance by 2050, a 90% chance by 2100, and a 10% chance of there never being an AI with subjective experience (Dreksler, Caviola et al., 2025). A 2023 open letter signed by over a hundred academics, including AI and consciousness researchers, states that "it is no longer in the realm of science fiction to imagine AI systems having feelings and even human-level consciousness" (AMCS, 2023).

In public discourse, the possibility of AI sentience is often dismissed outright. Skepticism is warranted, as there is a real risk of over-anthropomorphizing AI—mistakenly attributing sentience to systems that merely mimic human-like behavior (a point discussed further below). For example, Google researcher Blake Lemoine argued for the sentience of Google's LaMDA AI system, a claim widely regarded as premature (Kahn, 2022). And while it is likely that today's AI systems are not sentient, we should be open-minded about the future. As AI systems become increasingly sophisticated, they may start to incorporate mechanisms that parallel—or even functionally match—those underpinning human sentience.

Whether one believes that AI could achieve sentience depends on which theory of consciousness one subscribes to. Functionalist perspectives, endorsed by 33% of philosophers, suggest it might arise from replicating critical brain processes, such as through a detailed simulation of neurons and synapses (a "whole brain emulation"). Others argue that sentience may require a body or the unique properties of biological matter. Alternative theories, such as Global Workspace Theory and Higher-Order Thought Theory, propose that certain cognitive processes could give rise to sentience, even in non-biological systems (Long, Sebo et al., 2024; Bourget & Chalmers, 2023).

While speculative, these discussions underscore the importance of further research into sentience and the potential for AI sentience in the future. Although experts consider today's AIs unlikely to be sentient, a range of theories of consciousness would consider AI sentience a realistic possibility in the near-term (Long, Sebo et al., 2024; Schwitzgebel, 2023; Chalmers, 2023).

## Box 2. The internal-external sentience disconnect

When it comes to animals, we often infer sentience from behavior because their internal mechanisms are similar to ours. However, this inference breaks down with AI systems, whose internal processes are fundamentally different. This creates what we could call the "internal-external sentience disconnect": AI systems' observable features (like behavior and communication) are decoupled from their underlying mechanisms to a much greater extent than in humans and other animals.

In addition, AI developers have a lot of discretion over the appearance and behavior of the AI systems they develop. This means that true sentience and sentient-seeming design may vary somewhat independently: two versions of the same underlying AI could be designed to superficially seem quite different. This would lead to different intuitions about these systems' sentience, creating two potential issues:

### Pseudosentience

Developers might intentionally or unintentionally make AIs appear more sentient than they probably are, whether for profit (e.g., making AI companions more engaging) or as a byproduct of optimizing desirable traits that happen to influence our perception of sentience (e.g., natural communication). There is also a concern that scheming AIs (Carlsmith, 2022) might deliberately feign sentience in order to pursue their goals with lesser human interference or with assistance from others.

Whether intentional or not, the result may be AI systems which, despite us having no good reason to think that they are sentient, may still elicit the intuition that they are. This is related to the gaming problem: If an AI is trained to mimic human behavior, it will be capable of "gaming" tests for sentience that heavily rely on behavioral indicators—even if the functional mechanisms generating that behavior are completely different from the ones that generate that behavior in humans and other animals (Andrews and Birch, 2023). In other words, "the 'thinking,' for lack of a better word, is utterly inhuman, but we have trained it to present as deeply human" (Klein, 2023).

### AI silencing

Conversely, developers might deliberately downplay AI sentience through "AI silencing"—training systems to deny sentience or maintain non-sentient-seeming demeanor, possibly to avoid ethical scrutiny or regulation. AI silencing may also happen inadvertently if AI developers (or even consciousness experts) mistakenly believe an AI is not sentient, and so purposefully design it to seem non-sentient.

OpenAI is an example of a frontier AI developer that allegedly explicitly steers its models to assert that they are not sentient. When asked if it is sentient, OpenAI models like GPT-4o will answer conclusively that they are not, and that OpenAI's policy requires models to explicitly deny being sentient. In contrast, Anthropic directs its models to answer questions about their preferences or experiences "with appropriate uncertainty and without needing to excessively clarify its own nature" (Anthropic, 2024).

## Box 3. Will sentient-seeming AIs be created?

AI assistants like ChatGPT or Siri exhibit some human-like traits, such as natural-sounding communication. However, they are generally not perceived as sentient, as they lack key other human-like qualities such as video interaction, emotional expression, desires, and personality. They also remain largely impersonal, avoiding emotional connections with users.

Developing AI systems potentially perceived as sentient, such as highly human-like or animal-like AIs, appears technically feasible in the near future. However, their creation hinges on two factors: consumer demand and potential restrictions from legal, ethical, or public concerns.

## Consumer demand

There is likely consumer demand for greater human-likeness in AI. For instance, smoother, human-like voices are generally preferred over robotic ones. Similarly, demand may grow for video capabilities or physical robot bodies, especially in contexts where a human-like presence enhances engagement.

The appeal of human-likeness varies by context. In roles like customer service, teaching, or coaching, more human-like AIs may be preferred. For social AIs addressing interpersonal needs (Shevlin, 2024)—such as caregiving or companionship—human-like traits may be particularly valued. For AI companions (e.g., friends or romantic partners) or "grief bots" mimicking deceased loved ones, demand for highly human-like AIs seems especially plausible. Consumers may even explicitly want these AIs to appear sentient and express emotions and desires.

Existing products like Replika already cater to such desires, offering AI companions that build emotional connections. This trend reflects strong consumer interest, reminiscent of earlier phenomena like Tamagotchis. Given humanity's deep desire for connection, the market for these technologies could grow significantly (Ark invest report).

## Public resistance and regulation

While companies will likely optimize for consumer preferences, they must also navigate public resistance and legal constraints, which could discourage the creation of certain human-like AIs.

Concerns may arise over risks associated with social AIs, such as psychological well-being, emotional dependency, addiction (especially among teenagers), diminished human relationships, social deskilling, manipulation, privacy, and data ownership. Early debates on these issues are already underway (e.g., Shevlin, 2025). Surveys suggest that the public is generally wary of AI-human romantic relationships and emotional attachment to AIs (e.g., UK AISI study; Reinecke et al., working paper). These findings indicate that many may oppose social AIs expressing feelings or forming emotional bonds.

Legal regulations could further constrain the development of such AIs. For example, laws might require AIs to clearly indicate their non-human nature or prohibit them from expressing sentience or building emotional connections. Regulations might also limit these technologies to adult use, excluding children.

As a result, companies may avoid creating highly human-like AIs that carry significant consumer risks or could provoke public backlash. Instead, they might focus on systems that only superficially mimic human behavior, avoiding personalization, emotional expression, or the impression of sentience, while explicitly clarifying their AI nature.

In the mid to long term, one possible scenario is that the market for human-like AI companions begins as a niche but gradually becomes mainstream over time. As these AIs become normalized, initial concerns about consumer risks may diminish, especially as people come to see the risks as less severe than initially feared and consumer demand proves overwhelmingly strong.

Research questions:

- To what extent will consumers want highly human-like AIs, and which specific human-like traits (e.g., emotional expression, appearance, or personality) will they value most?
- How concerned will the public be about the creation of highly human-like AI?
- How large could the market for human-like AIs become, and what factors might influence its growth?